\documentclass[reprint,amsmath,amssymb,aps,superscriptaddress]{revtex4-2}

\usepackage{graphicx}
\usepackage{dcolumn}
\usepackage{amsmath}
\usepackage{bm}
\usepackage{mathrsfs}
\usepackage{txfonts}
\usepackage{epstopdf}
\usepackage{subfigure}
\usepackage{caption}
\usepackage{float}

\begin{document}
	
\title{Strain modulation of photocurrent in Weyl semimetal TaIrTe$_4$}
	
\author{Ying Ding}
\affiliation{Department of Physics, Hangzhou Dianzi University, Hangzhou, Zhejiang 310018, China}
\author{Xinru Wang}
\affiliation{Department of Physics, Hangzhou Dianzi University, Hangzhou, Zhejiang 310018, China}
\author{Liehong Liao}
\affiliation{Department of Physics, Hangzhou Dianzi University, Hangzhou, Zhejiang 310018, China}
\author{Xinyu Cheng}
\affiliation{Department of Physics, Hangzhou Dianzi University, Hangzhou, Zhejiang 310018, China}
\author{Jiayan Zhang}
\affiliation{Department of Physics, Hangzhou Dianzi University, Hangzhou, Zhejiang 310018, China}
\author{Yueyue Wang}
\affiliation{Department of Physics, Hangzhou Dianzi University, Hangzhou, Zhejiang 310018, China}
\author{Hao Ying}
\email{yinghao@hdu.edu.cn}
\affiliation{Department of Physics, Hangzhou Dianzi University, Hangzhou, Zhejiang 310018, China}
\author{Yuan Li}
\email{liyuan@hdu.edu.cn}
\affiliation{Department of Physics, Hangzhou Dianzi University, Hangzhou, Zhejiang 310018, China}
\email[]{Corresponding author: liyuan@hdu.edu.cn}
	
\date{\today}
\begin{abstract}
We study the effect of the strain on the energy bands of TaIrTe$_4$ sheet and the photocurrent in the Cu-TaIrTe$_4$-Cu heterojunction by using
the quantum transport simulations. It is found that the Weyl points can be completely broken with increasing of the strain along z dirction. One can obtain a large photocurrent in the Cu-TaIrTe4-Cu heterojunction in the absence of the strain. While the photocurrent can be sharply enhanced by the strain and reach
a large value. Accordingly, the maximum values of the photocurrent can be explained in terms of the transitions between peaks of density of states and band structures. The strain-induced energy bands and photocurrent exhibit anisotropic behaviors.
Our results provide a novel route to effectively modulate the energy bands and the photocurrent by utilizing mechanical methods for TaIrTe$_4$-based devices.
\end{abstract}
\maketitle

Weyl semimetal (WSM) is a kind of topological state
of matter with non-trivial band structure, and is one of hot topics in lately condensed matter reseasrches. According to band characteristics, WSMs can be divided into two types. Type-I WSM is represented by TaAs~\cite{TaAs}, whose band near the Weyl point is an upright X cone, and fermions satisfy Lorentz transformation symmetry. The characteristic of type-II WSM is that the Dirac cone near the Weyl point is tilted, so the corresponding electron dispersion relation near the Weyl point does not satisfy the Lorentz  symmetry. After WTe$_2$~\cite{WTe2}and MoTe$_2$~\cite{MoTe2} were proved to be type-II WSMs, TaIrTe$_4$ was also predicted as a type-II WSM. TaIrTe$_4$ has novel features such as it hosts only four well-separated Weyl points~\cite{origin}, which is the minimum number of Weyl points for a system with time-reversal invariance. It was further noted that the band structure~\cite{band} and noncentrosymmetric Weyl states~\cite{exp} in TaIrTe$_4$ are accessed by using angle-resolved photoemission spectroscopy and the topological Fermi arcs can be found. Strong-coupling anisotropic s-wave superconductivity~\cite{superconductivity,surface supercoductivity} and quantum spin Hall effect~\cite{Halleffect} have been discovered in TaIrTe$_4$. Moreover, nonlinear photoresponse~\cite{Nonlinear} and broadband anisotropic photoresponse of the hydrogen atom version~\cite{hydrogen atom} of TaIrTe$_4$ have been investigated. It has also been demonstrated that room temperature nonlinear Hall effect and wireless radio frequency rectification occur in the Weyl semimetal TaIrTe$_4$~\cite{room}. The strain can induce electric currents due to the chiral magnetic effect in WSMs and surface plasmon polaritons have been reported in strained WSMs~\cite{strainWeyl}. Mechanical strain can be used to obtain large enhancement of photogalvanic effects in the phosphorene-based photodetector~\cite{strain} and enhance photogalvanic effect in MgCl$_2/$ZnBr$_2$ heterojunction~\cite{MgCl2/ZnBr2}. The strain can modulate the valley and charge transport of a WSM-based n-p-n junction~\cite{npn}. It is reported that WSMs have the possibility of implementing a pseudo-magnetic filed which can be large enough to form quantized Landau levels with small amplitude of the strain~\cite{magnetic}. However, the study on TaIrTe$_4$-based device under the illumination of linearly polarized light is still scarce. Therefore, the properties of TaIrTe$_4$ with linearly polarized light and how to regulate it deserve to further study. In this paper, we propose an effective strain modulation of the photocurrent induced by the linearly polarized light in the TaIrTe$_4$ heterojunction.
\begin{figure}[b]
	\centering
	\includegraphics[scale=0.15]{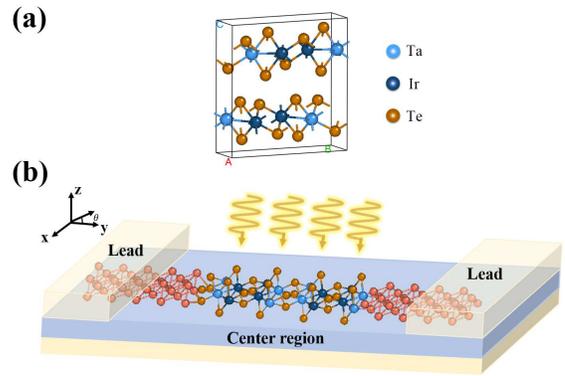}
	\caption{(a) The unit cell structure of TaIrTe$_4$, (b) Schematic of the Cu-TaIrTe$_4$-Cu transport model. The central part of TaIrTe$_4$ heterojunction is irradiated by linearly polarized light. }
	\label{fig:model}
\end{figure}

As shown in Fig~\ref{fig:model}(a), the space group of TaIrTe$_4$ is Pmn2$_1$ and  the lattice constants are $a=3.77{\AA}$, $b=12.421{\AA} $, $c=13.184{\AA}$, which are consistent with those given in previous studies~\cite{band,band1}. The cutoff energy is set to be $100$ Hartree (1 Hartree$= 27.21 \mathrm{eV}$). Brillouin zone integration is performed with a 15 $\times$ 7$\times$ 7 k-point mesh for geometry optimization and self-consistent electronic structure calculations.

We choose Cu as electrodes of the TaIrTe$_4$ heterojunction because Cu matches well with TaIrTe$_4$. The transport direction is along y axis, and the vacuum direction is along z axis. Note that the transverse width of the heterojunction is infinite. The exchange-correlation potentials are approximated by using the generalized gradient approximation (GGA) functional as parameterized by Perdew, Burke and Ernzerhof (PBE).
When the central region is irradiated by the linearly polarized light, the photogenerated current can flow from electrodes into the central region. Based on the linear response approximation, the photogenerated current can be obtained~\cite{phc0,phc}:
\begin{eqnarray}\label{eq:parameter}
\begin{aligned}
J_{L}^{\left ( ph\right )}&=\frac{ie}{h}\int  \left \{
\cos^{2}\theta \mathrm{Tr}\left \{\Gamma _{L}\left [ G_{1}^{<\left ( ph\right )}+f_{L}\left ( G_{1}^{>\left ( ph\right )}-G_{1}^{<\left ( ph\right )}\right )\right ]\right \} \right.\\
&+ \sin^{2}\theta \mathrm{Tr}\left \{\Gamma _{L}\left [ G_{2}^{<\left ( ph\right )}+f_{L}\left ( G_{2}^{>\left ( ph\right )}-G_{2}^{<\left ( ph\right )}\right )\right ]\right \}  \\
 &+2 \sin\left ( 2\theta \right ) \mathrm{Tr}\left \{\Gamma _{L}\left [ G_{3}^{<\left ( ph\right )}+f_{L}\left ( G_{3}^{>\left ( ph\right )}-G_{3}^{<\left ( ph\right )}\right )\right ]\right \} \Big \}\ \mathrm{d}E,
\end{aligned}
\end{eqnarray}
where
\begin{eqnarray}
\begin{array}{l}
 G_1^{ > ( < )ph} = \sum\limits_{\alpha ,\beta  = x,y,z} {{C_0}NG_0^r} {e_{1\alpha }}p_\alpha ^ \dag G_0^{ > ( < )}{e_{1\beta }}{p_\beta }G_0^a, \\
 G_2^{ > ( < )ph} = \sum\limits_{\alpha ,\beta  = x,y,z} {{C_0}NG_0^r} {e_{2\alpha }}p_\alpha ^ \dag G_0^{ > ( < )}{e_{2\beta }}{p_\beta }G_0^a, \\
 G_3^{ > ( < )ph} = \sum\limits_{\alpha ,\beta  = x,y,z} { {{C_0}} NG_0^r({e_{1\alpha }}p_\alpha ^ \dag G_0^{ > ( < )}{e_{2\beta }}{p_\beta }}\\
 \hspace{2cm}+{{e_{2\alpha }}p_\alpha ^ \dag G_0^{ > ( < )}{e_{1\beta }}{p_\beta })G_0^a}.
 \end{array}
\end{eqnarray}
\begin{figure}[t]
	\centering
	\includegraphics[scale=0.08]{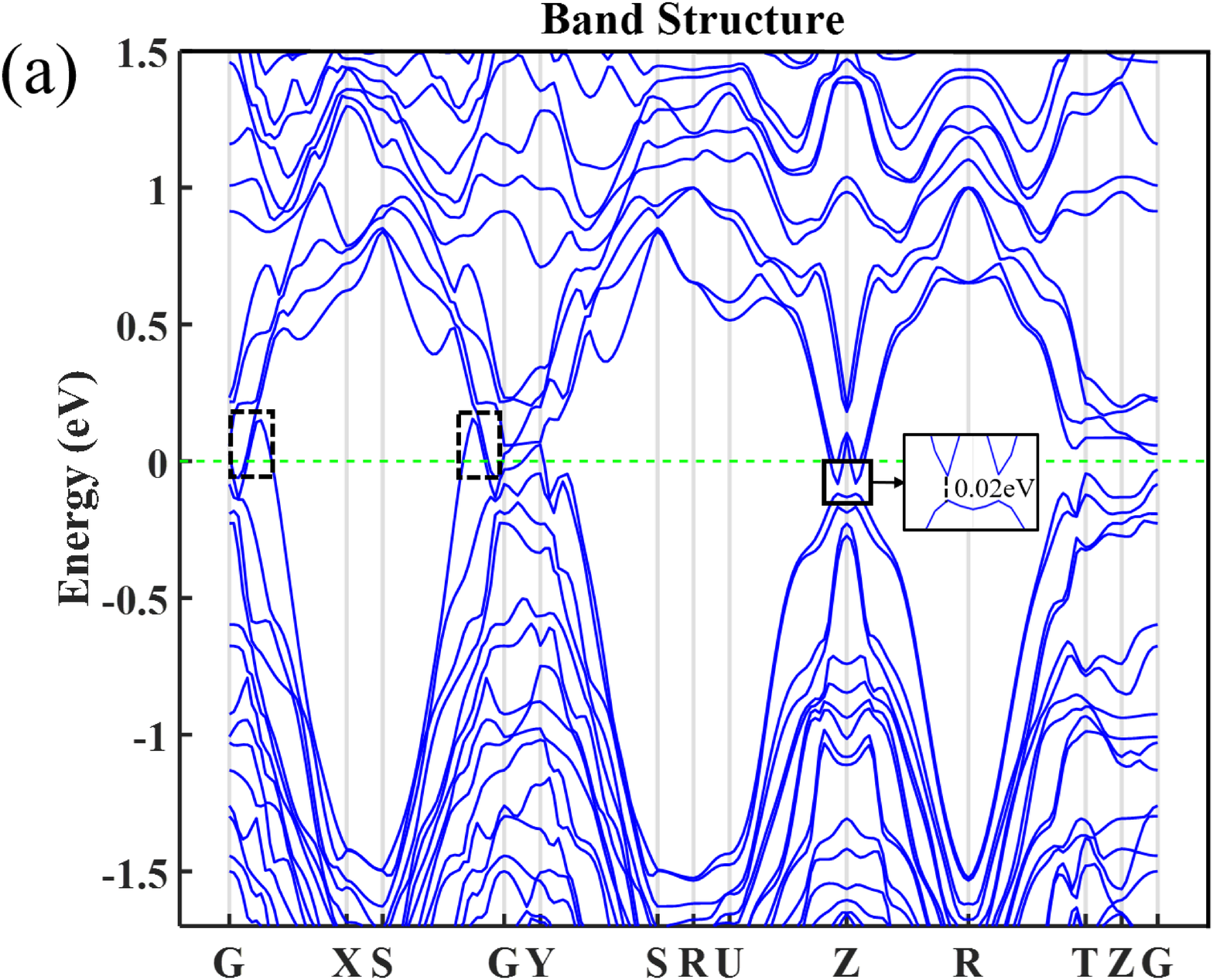}
	\includegraphics[scale=0.08]{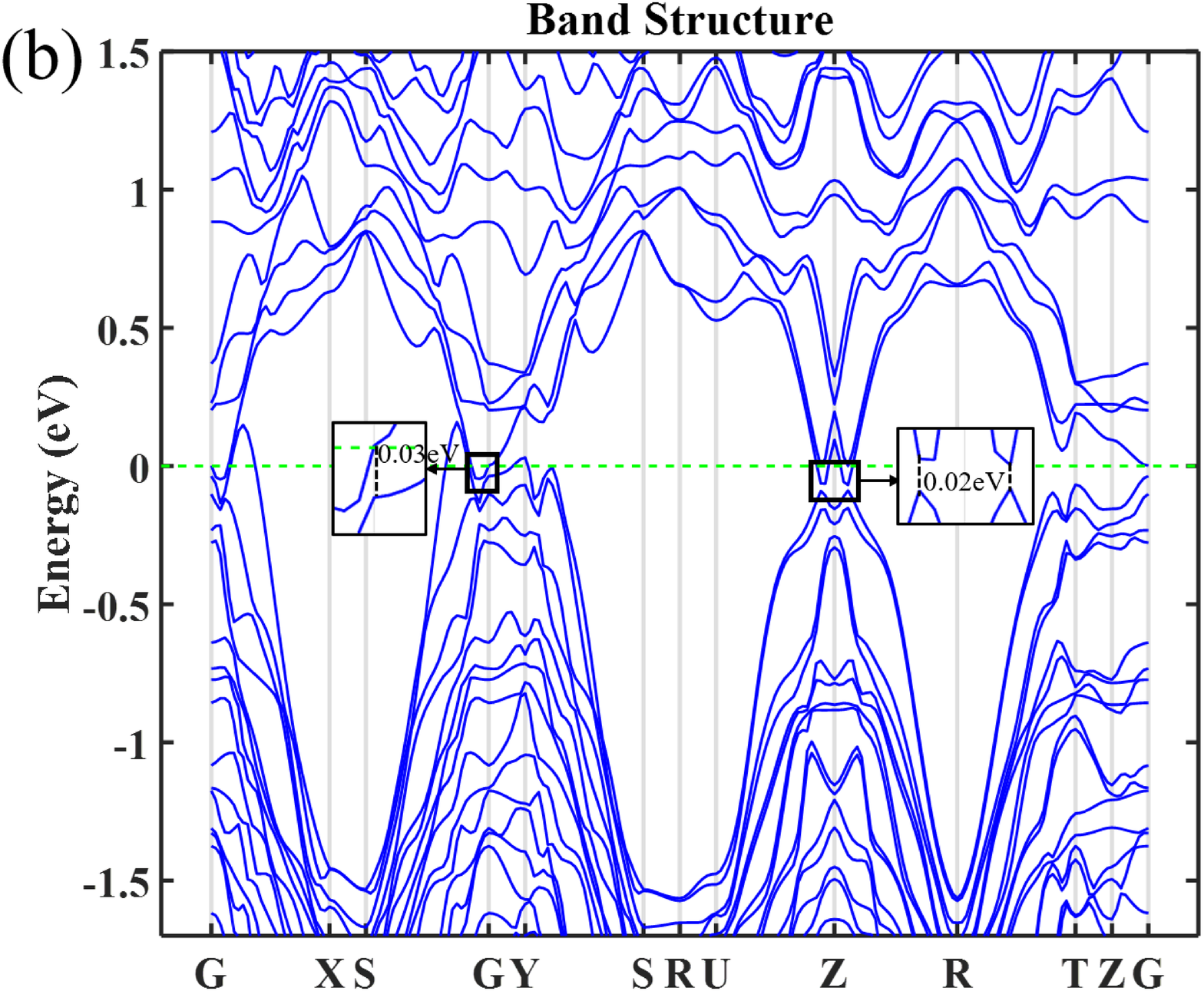}

    \includegraphics[scale=0.08]{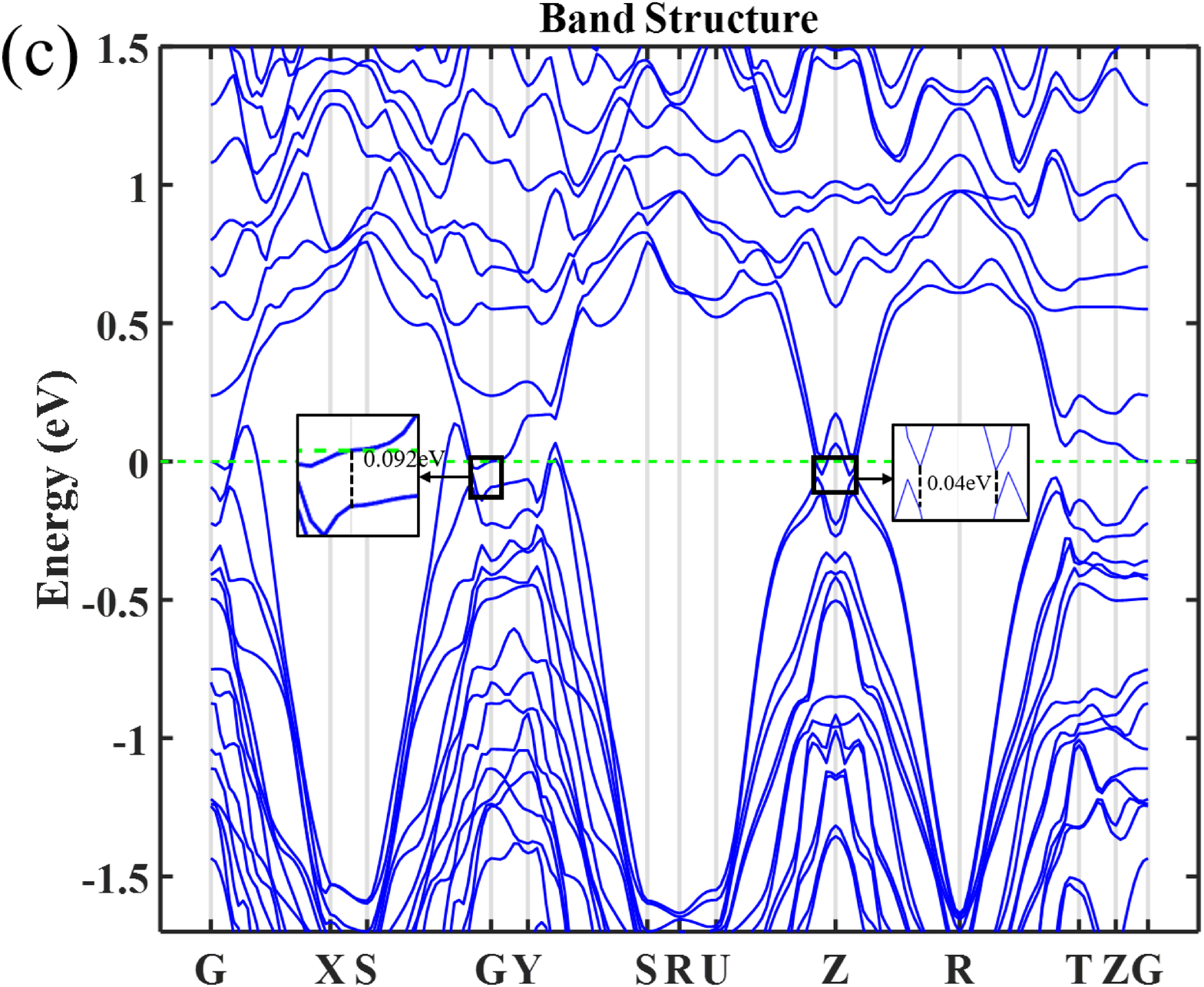}
	\includegraphics[scale=0.08]{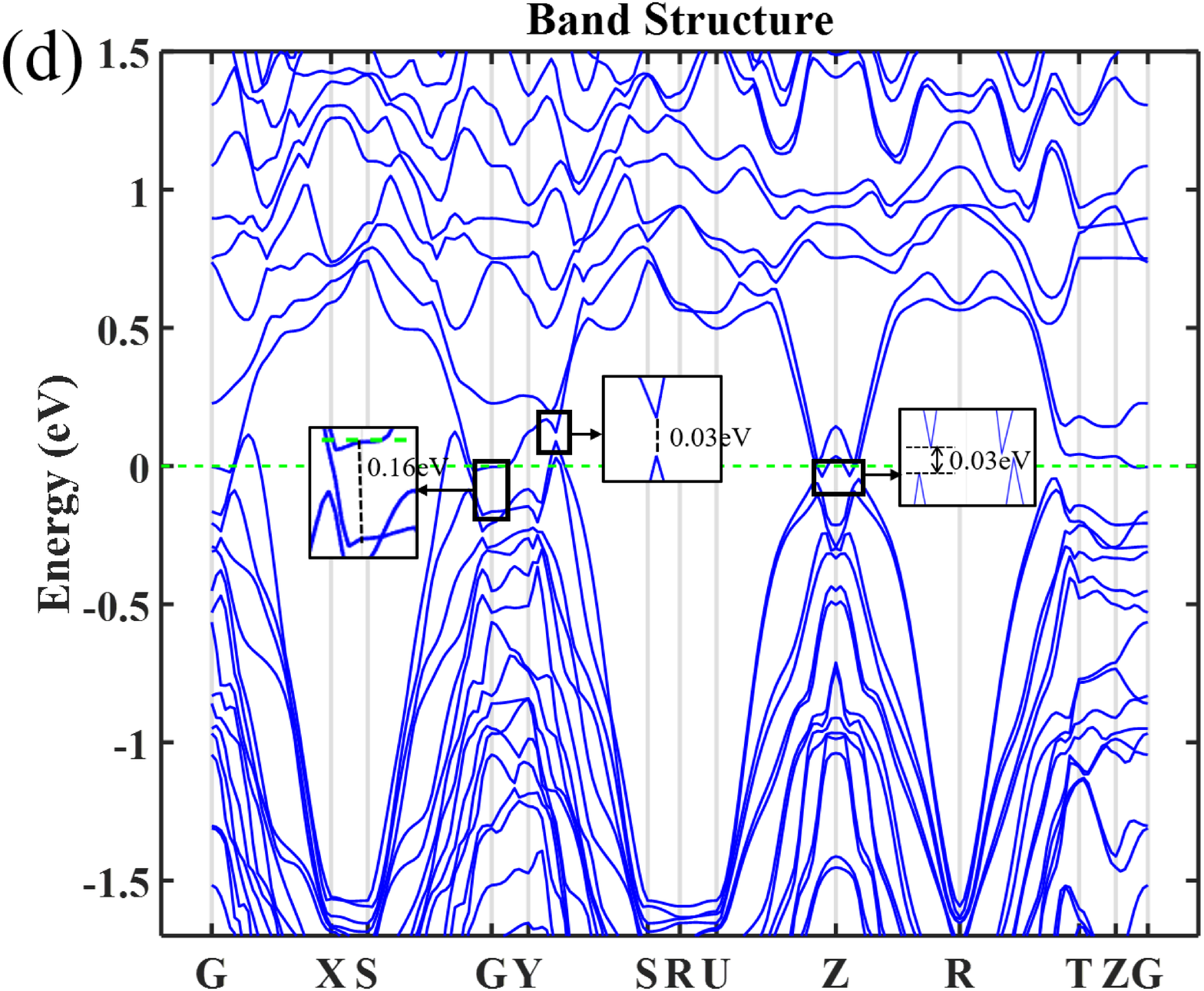}
	\caption{Energy band structure of TaIrTe$_4$ for different strain strengths exerted along x axis: (a)$\zeta =0  $, (b)$\zeta = 2\%  $ , (c)$\zeta = 6\%  $ , (d)$\zeta = 8\%. $ }
	\label{fig:xB}
\end{figure}

Where $G_{1,2,3}^{ > ( < )ph}$ represent the greater and lesser Green's functions for electron-phonon interactions~\cite{phc yizhijie,phc2}, ${C_0} = \frac{{{e^2}\hbar \sqrt {{\mu _r}{\varepsilon _r}} }}{{2N{m^2}\omega \varepsilon c}}{I_\omega }$. Here $I_\omega$ is the number of photons per unit area and time, and $N$ is the number of photons. For linear polarized light, the polarization vector is $\hat e = \cos \theta {\hat e_1} + \sin \theta {\hat e_2}$,  $\theta$ indicating the angle between the direction of polarized light and the transport direction. The normalized photocurrent is given by
$J = J_L^{(ph)}/e{I_\omega }$.

\begin{figure}[t]
	\centering
	\includegraphics[scale=0.5]{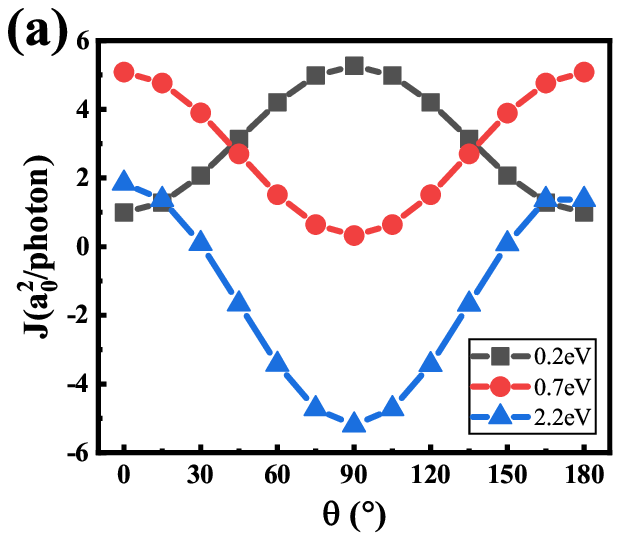}
	\includegraphics[scale=0.5]{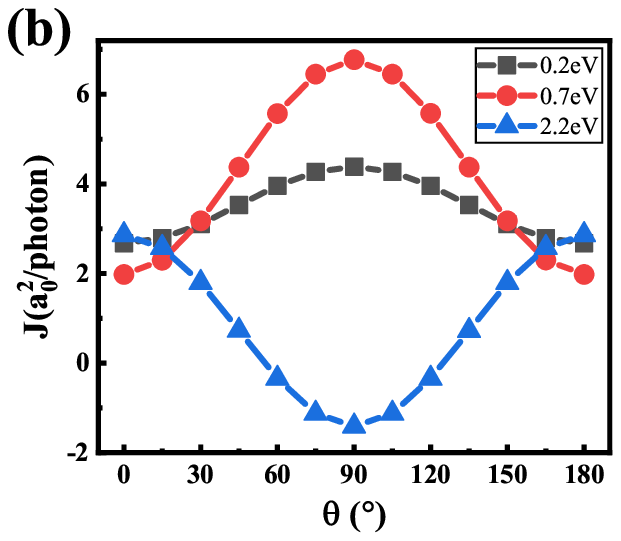}

    \includegraphics[scale=0.5]{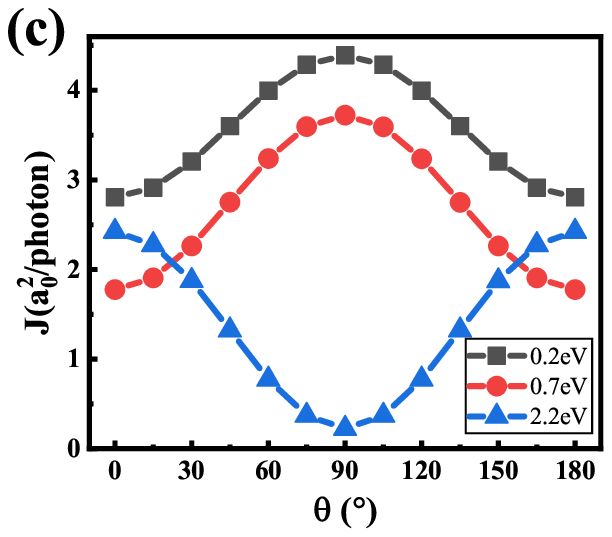}
	\includegraphics[scale=0.5]{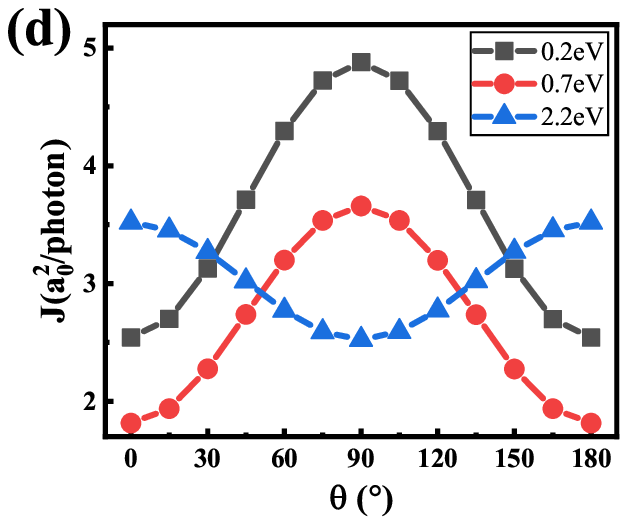}
	\caption{The photocurrent for  photon energies of 0.2eV, 0.7eV, and 2.2 eV under different strengths of the strain along x direction: (a) $\zeta = 0\% $, (b) $\zeta = 2\%$, (c) $\zeta = 6\% $, (d) $\zeta = 8\%. $ }
	\label{fig:cos}
\end{figure}
In our simulations, the strain is applied along two directions of TaIrTe$_4$, viz x and z direction. We can define the ratio  $\zeta=(b-b_{0})/b_{0}$ to denote the strength of strain, where $b$ and $b_{0}$ represent the lattice constants with and without strain, respectively. During the calculation, we control the strength of strain to be less than $10\%$ to ensure the authenticity of the TaIrTe$_4$ structure. Fig.~\ref{fig:xB} shows the energy band structure of TaIrTe$_4$ for exerting the strain along x direction.
When $\zeta=0$, we can obtain two Weyl points along the path $G - X$ and $S- G$ shown in the dotted boxes [see Fig.~\ref{fig:xB}(a)], which are in accord with the characteristics of the type-II Weyl semimetal~\cite{origin}. In addition, it can be found in the inset that there is a distinct band gap of 0.02 eV near the high-symmetry point Z.
Seen from Fig.~\ref{fig:xB}(b), when $\zeta=2\%$, there is an energy separation of 0.03 eV at the high-symmetry point G and the band gap keeps unchanged. When the strain increases to $6\%$, the band separation expands from 0.03 eV to 0.092 eV at the point G, and the band gap enlarges from 0.02 eV to 0.04 eV near the point Z. When the strain comes to $8\%$, it can be shown from Fig.~\ref{fig:xB}(d) that the energy separation at the point G is 0.16 eV, and the band gap near point Z has a relative shift to become an indirect band gap. Moreover, an energy separation of 0.03 eV appears between the path Y-S. Interestingly, the Weyl points remain intact and do not separate when different strains are applied on the central region along x direction. It can provide an indirect verification of the Weyl semimetal TaIrTe$_4$ by utilizing optical transitions between the strain-induced energy subbands.

In order to understand the effect of the strain on the optical transitions, we calculate the photocurrent of Cu-TaIrTe$_4$-Cu heterojunction in the presence and absence of the strain by using density function theory (DFT) combined with the non-equilibrium Green's function method (NEGF), which is executed in the quantum transport package $NanoDcal$~\cite{Nanodcal,Nanodcal2}. The photon energies change from 0.01eV to 3.0eV, which cover the most of infrared-light and visible-light range. Fig.~\ref{fig:cos} demonstrates the angular dependence of the photocurrent for different strengths of the strain applied along x direction. The photocurrent $J$ is found to exist a cosinusoidal dependence on the polarization angle $\theta$ for all photon energies and is governed by the $C_s$ symmetry, which is highly consistent with the phenomenological theory~\cite{CS,CS2}. It can be observed in Fig.~\ref{fig:cos} that the maximum photocurrent $J_{max}$ is acquired at either $\theta=0^{\circ}$ or $\theta=90^{\circ}$ for all photon energies. We next analyse the effect of the strain applied along x direction on the maximum value of the photocurrent.
\begin{figure}[t]
	\centering
	\includegraphics[scale=0.55]{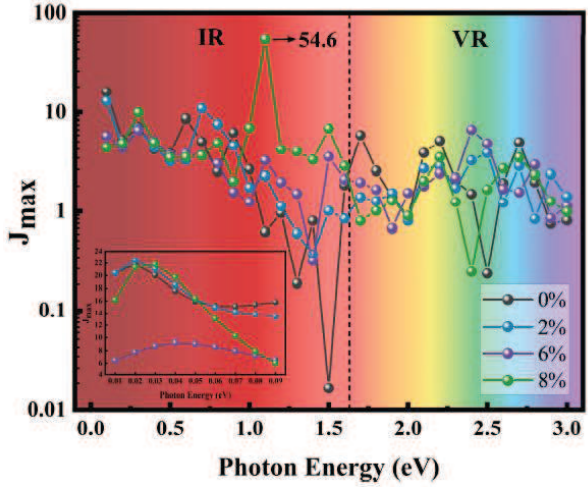}%
	\caption{The variation of the maximum photocurrent as a function of the photon energy for the strain exerted along x direction. }
	\label{fig:xphc}
\end{figure}

Fig.~\ref{fig:xphc} manifests $J_{max}$  as a function of photon energies changing from 0.1eV to 3.0 eV under varying strength of the strain.
The left side of dotted line is infrared-light region (IR) and the right side is visible-light region (VR).
The inset shows $J_{max}$ for the photon energy less than 0.1 eV. The photocurrent achieves a maximum value of 21.86 when the photon energyis about 0.02eV for $\zeta=0$. For the case $\zeta=0$, the photocurrent has six peaks in the photon energy range from 0.1 eV to 3.0 eV, which are 16.13 at 0.1eV, 8.63 at 0.6 eV, 6.28 at 0.9eV, 5.91 at 1.7 eV, 5.20 at 2.2 eV and 5.01 at 2.7eV. It means that one can obtain a large photocurrent in the Cu-TaIrTe$_4$-Cu heterojunction without strain. When the strain increases to $2\%$, the photocurrent obtains the largest value of 22.48 at 0.02 eV, which is in greatly agreement with the band gap near the point Z in Fig.~\ref{fig:xB}(b). There has two maximum values at 0.7 eV and 2.5 eV. When $\zeta=6\%$, the maximum value shifts to the energy 0.04eV, which is consistent with the change of the band gap from 0.02 eV to 0.04 eV near the point Z in Fig.~\ref{fig:xB}(c). Especially, under the action of $8\%$ strain, the photocurrent has five peaks. The photocurrent is sharply enhanced and can reach the largest value of 54.6 at 1.1 eV, which is more than 86 times of the photocurrent at $\zeta=0$. Evidently, the maximum photocurrent of TaIrTe$_4$-based heterojunction can reach larger values comparing with that of CrI$_3$~\cite{CrI3}, WTe$_2$ ~\cite{WTe2phc} and MoTe$_2$~\cite{MoTe2phc}.

\begin{figure}[t]
	\centering
	\includegraphics[scale=0.2]{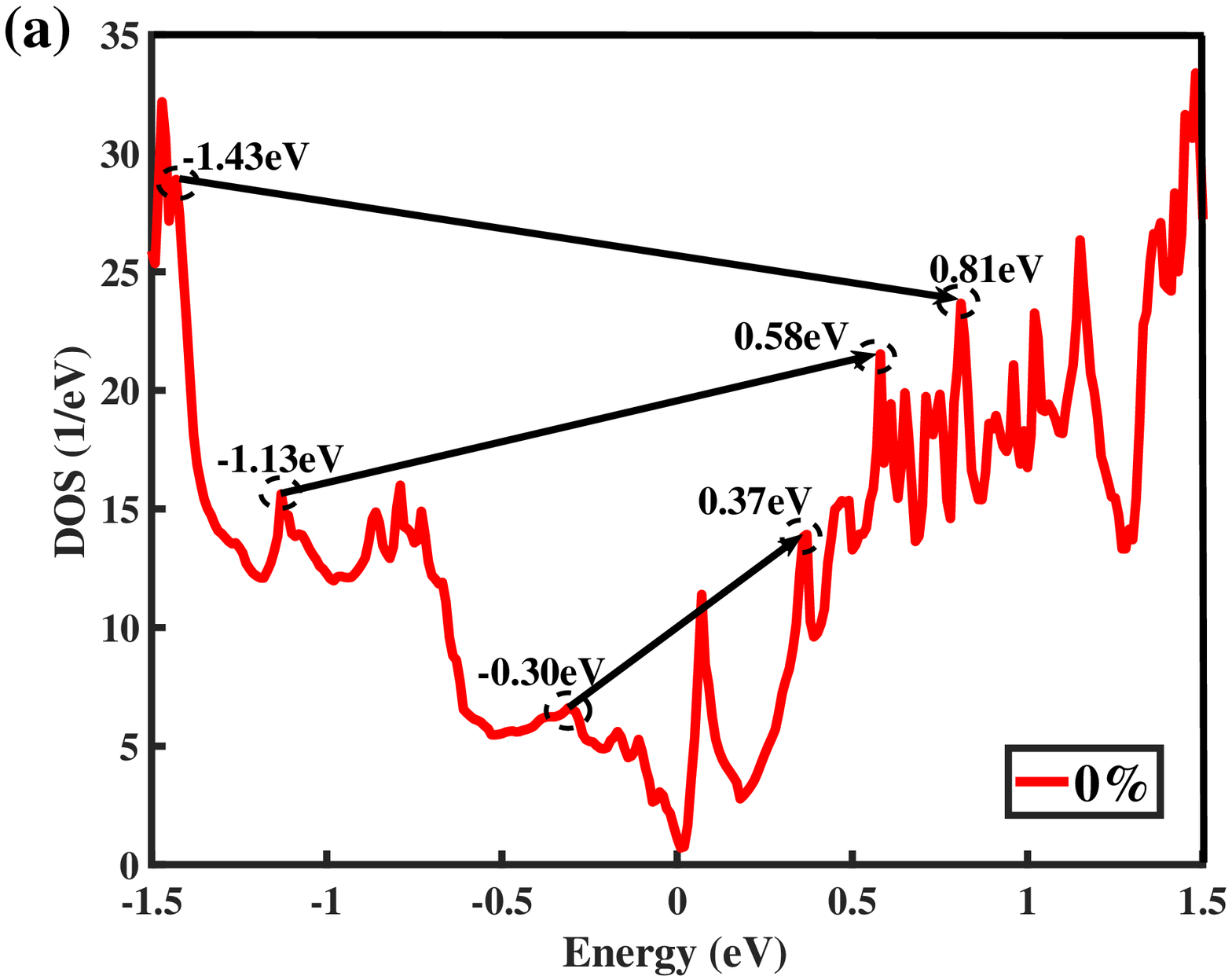}%
	\includegraphics[scale=0.2]{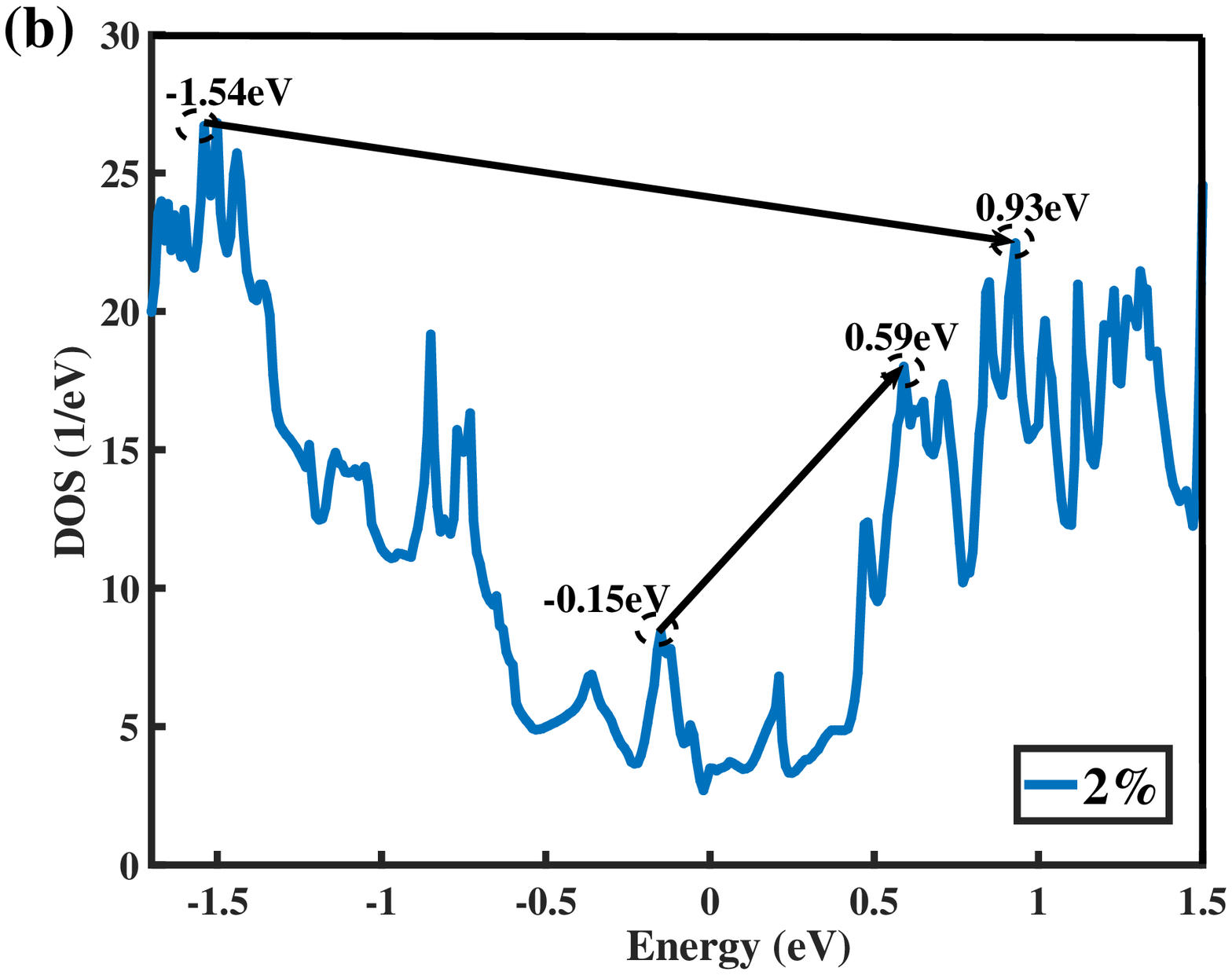}%

    \includegraphics[scale=0.2]{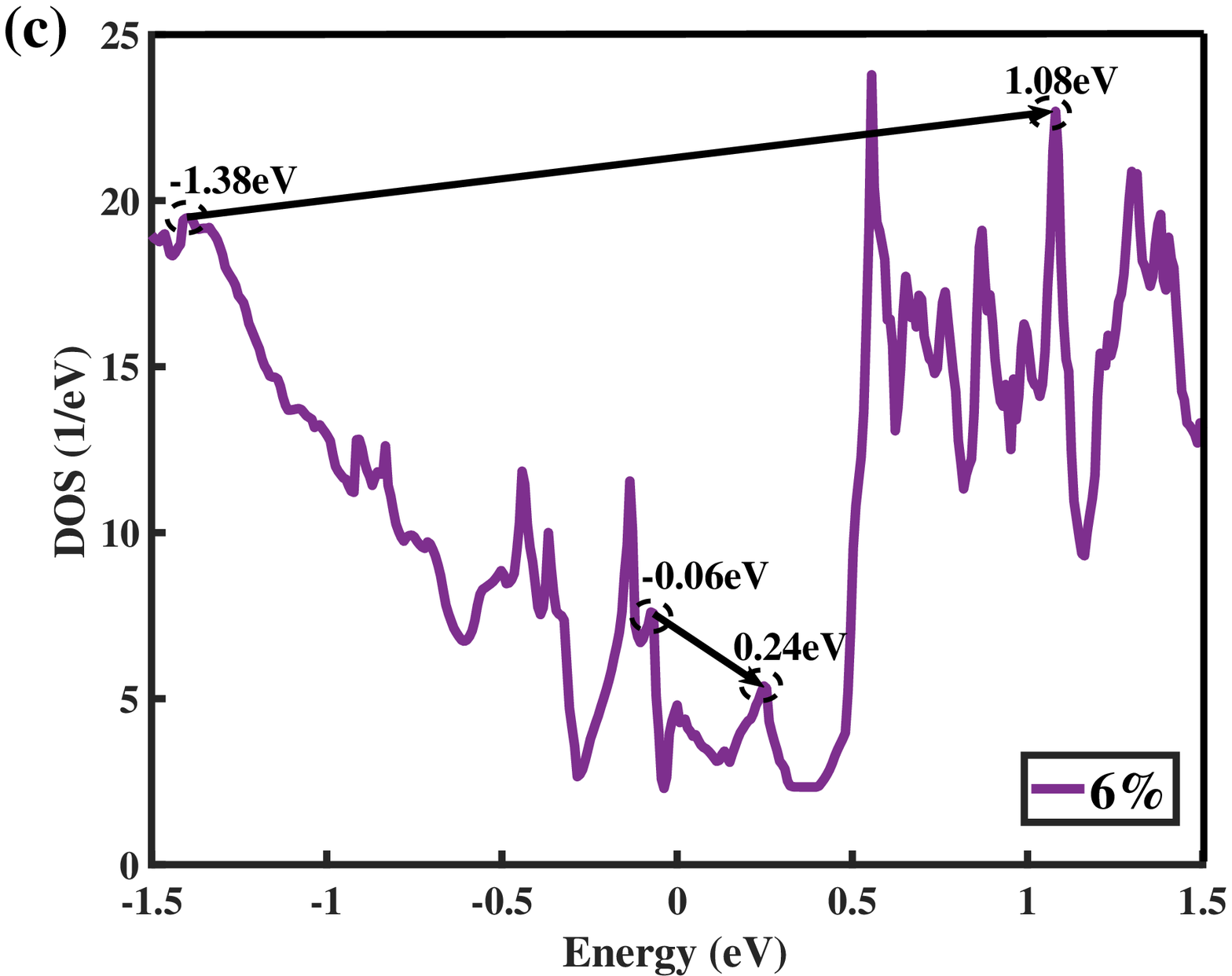}%
    \includegraphics[scale=0.2]{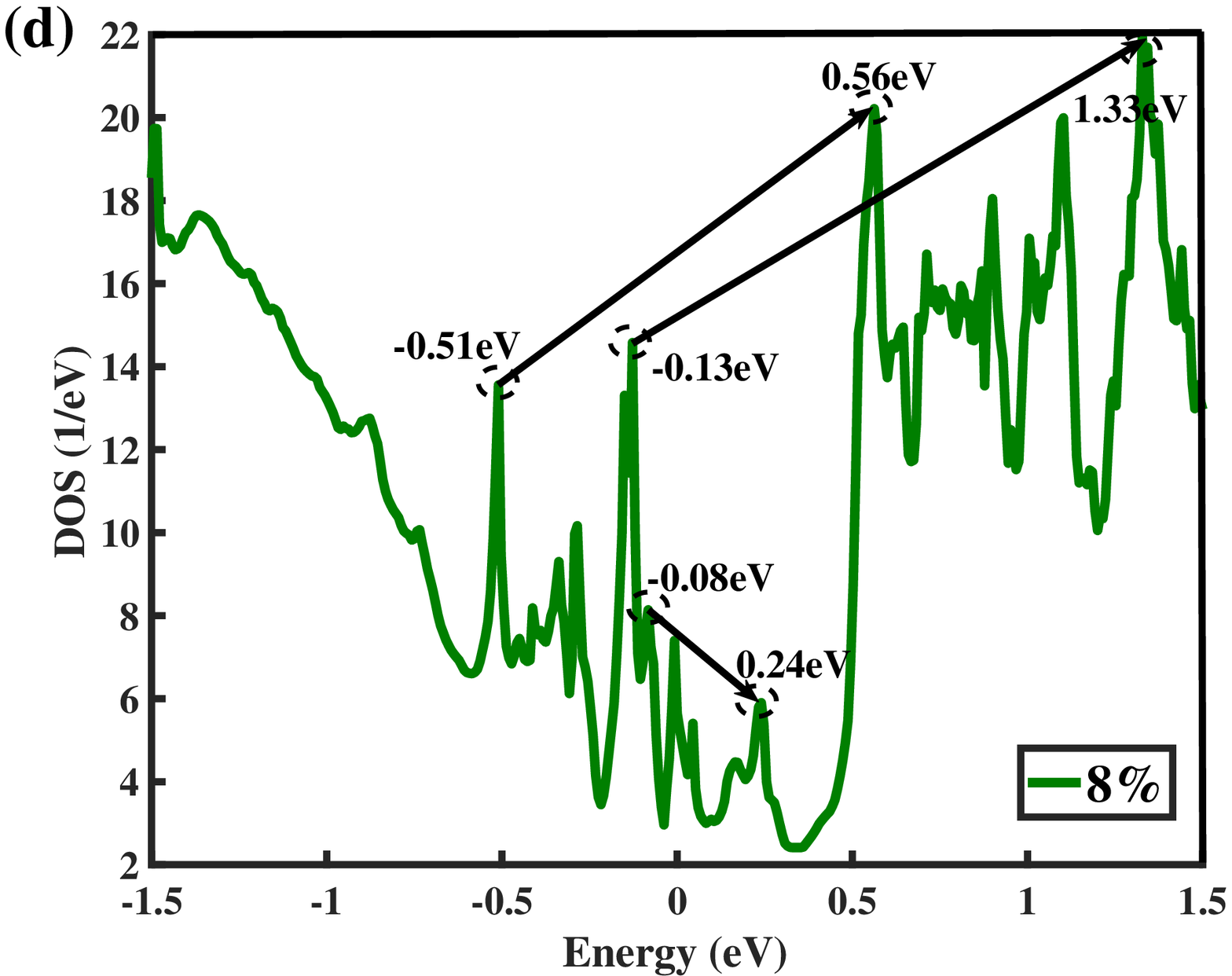}%
	\caption{Density of states of TaIrTe$_4$ for different strain strengths exerted along the x direction: (a)$\zeta = 0 $, (b)$\zeta = 2\%  $ , (c)$\zeta = 6\%  $ , (d)$\zeta = 8\% $.}
	\label{fig:xD}
\end{figure}

For further understanding the peaks of photocurrent, we plot the density of states (DOS) of TaIrTe$_4$, as demonstrated in Fig.~\ref{fig:xD}. According to Fermi's golden rule, the probability of transition for electrons between valence bands and conduction bands is proportional to DOS. Therefore, the electronic transitions between DOS peaks associate with large values of the photocurrent. As shown in Fig.~\ref{fig:xD}(a), there are several transitions [see black arrows], for example the transition from -1.13eV to 0.58eV associating with the large photocurrent at the photon energyof 1.7eV [see Fig.~\ref{fig:xphc}].
Similarly, the two transitions between peaks of DOS in Fig.~\ref{fig:xD}(b) can describe the large values of the photocurrent at the photon energies of 0.7eV and 2.5eV when $\zeta=2\%$. Obviously, the transition from -1.38eV to 1.08eV induces a large photocurrent around the photon energy2.4eV for $\zeta=6\%$.
Especially, when the strain is increased to $\zeta=8\%$, it can be presented in Fig.~\ref{fig:xD}(d) that the transition from -0.51eV to 0.56eV of the DOS peaks results in the largest photocurrent of 54.57 at 1.1eV. Basically, the maximum values of the photocurrent can be explained in terms of the transitions between DOS peaks and band structures.

\begin{figure}[t]
	\centering
	\includegraphics[scale=0.55]{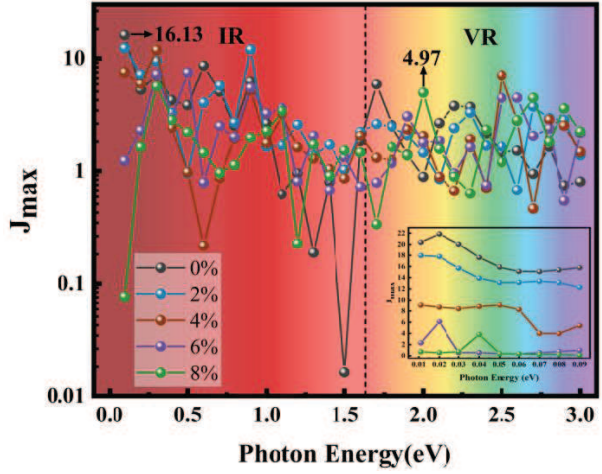}%
	\caption{The variation of the maximum photocurrent as a function of the photon energy for the strain exerted along z direction.}
	\label{fig:zphc}
\end{figure}

Next, we study the anisotropic behavior of the effect of the strain exerted along z direction. There exists a band gap about 0.02 eV near the high-symmetry point Z, which is slightly changed with increasing of the strain. Different from the strain configuration exerted along x direction, the Weyl points can be completely broken when the strength of the strain is larger than $8\%$.
The variations of the energy bands can be detected by measuring the strain-induced photocurrent. As shown in Fig.~\ref{fig:zphc}, when the photon energy is less than 0.1eV, the photocurrent significantly decreases from 21.86 to about 0.07 with increasing of the strain. It means that the strain-induced changes of the Weyl points can be measured by utilizing the photocurrent induced by the infrared light. When the strain is increased to $\zeta=2\%$, there exists four peaks of the photocurrent at the photon energies of 0.3eV, 0.9eV, 2.3eV and 2.7eV, and the maximum value of photocurrent is 12.01 at 0.9 eV. Moreover, it can be seen that the photocurrent reaches the maximum value of 11.88 at the photon energy of 0.3 eV when $\zeta=4\%$. An extra photocurrent peak occurs at the photon energy of 2.5eV. For the case $\zeta=6\%$, four photocurrent peaks occur at the photon energies of 0.3eV, 0.5eV, 0.9eV and 2.6eV. The maximum value of the photocurrent is 7.50 at 0.5eV. Differently, when $\zeta=8\%$, the photocurrent peaks have significant changes, which occur at the photon energies of 0.3eV, 1.1eV, 2.0eV and 2.7eV. Obviously, the TaIrTe$_4$ system can provide a platform with large photocurrents.
The strain can be utilized to modulate the photocurrent and detect the variation of energy bands of Weyl semimetals.

\begin{figure}[t]
	\centering
	\includegraphics[scale=0.2]{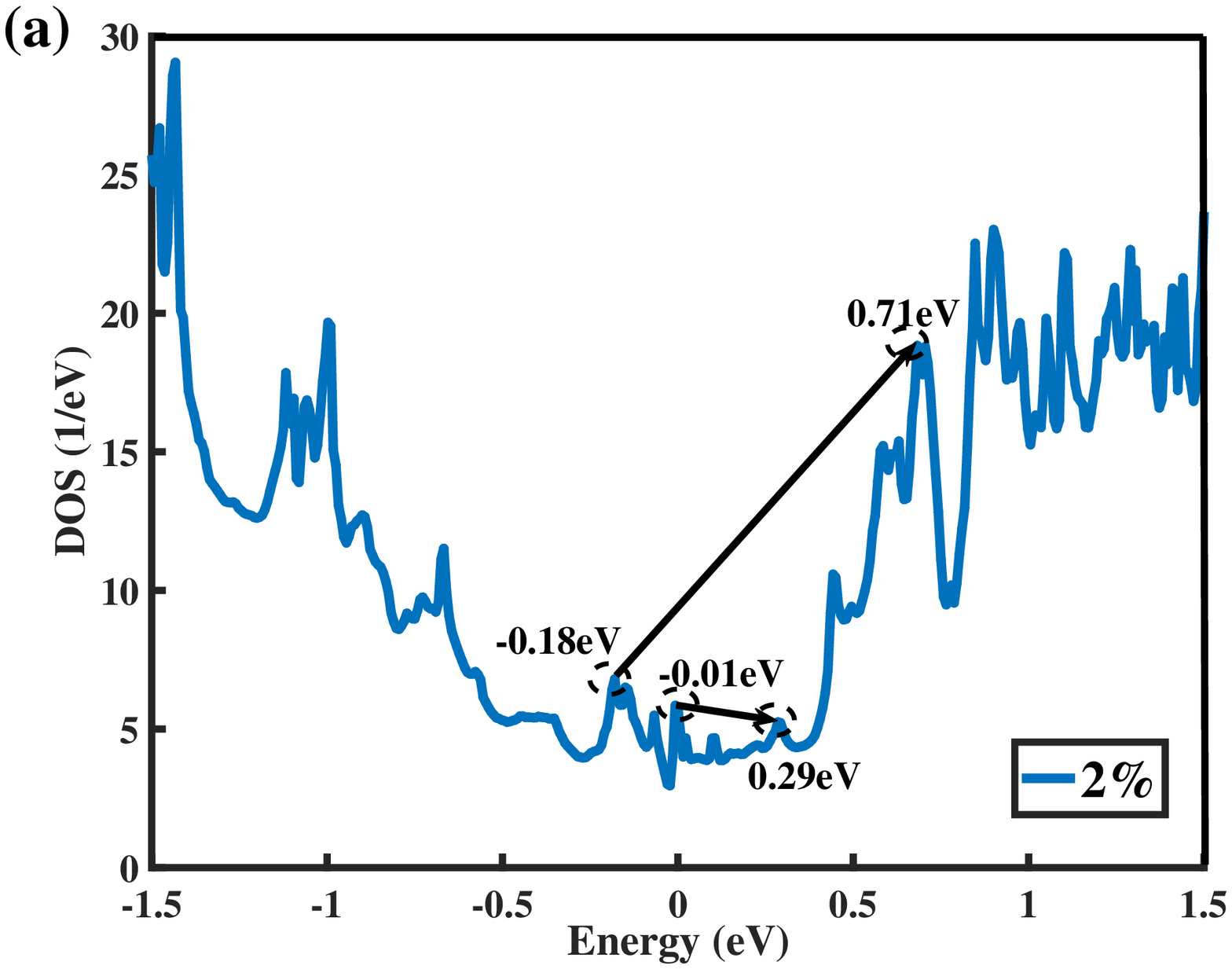}%
	\includegraphics[scale=0.2]{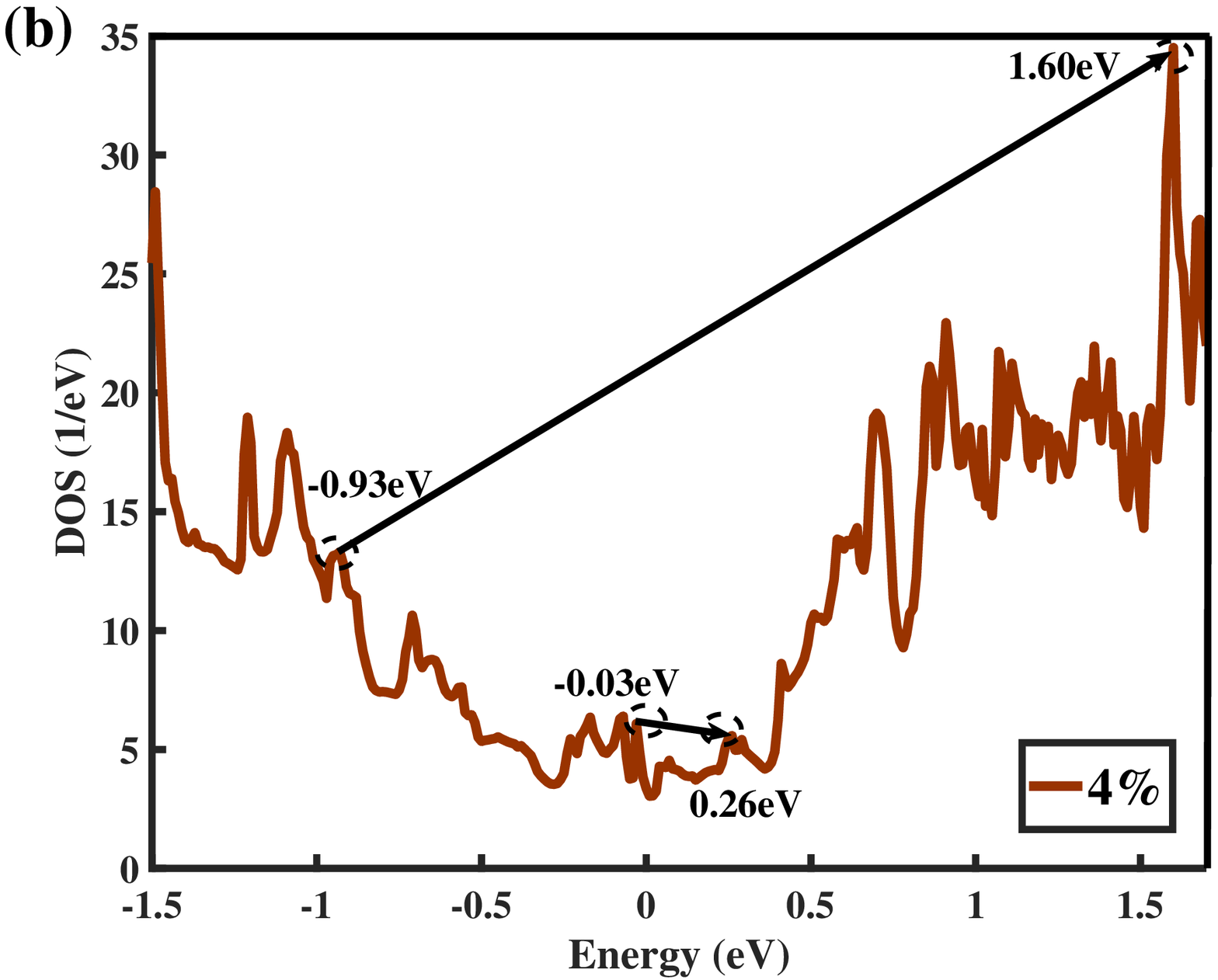}%

    \includegraphics[scale=0.2]{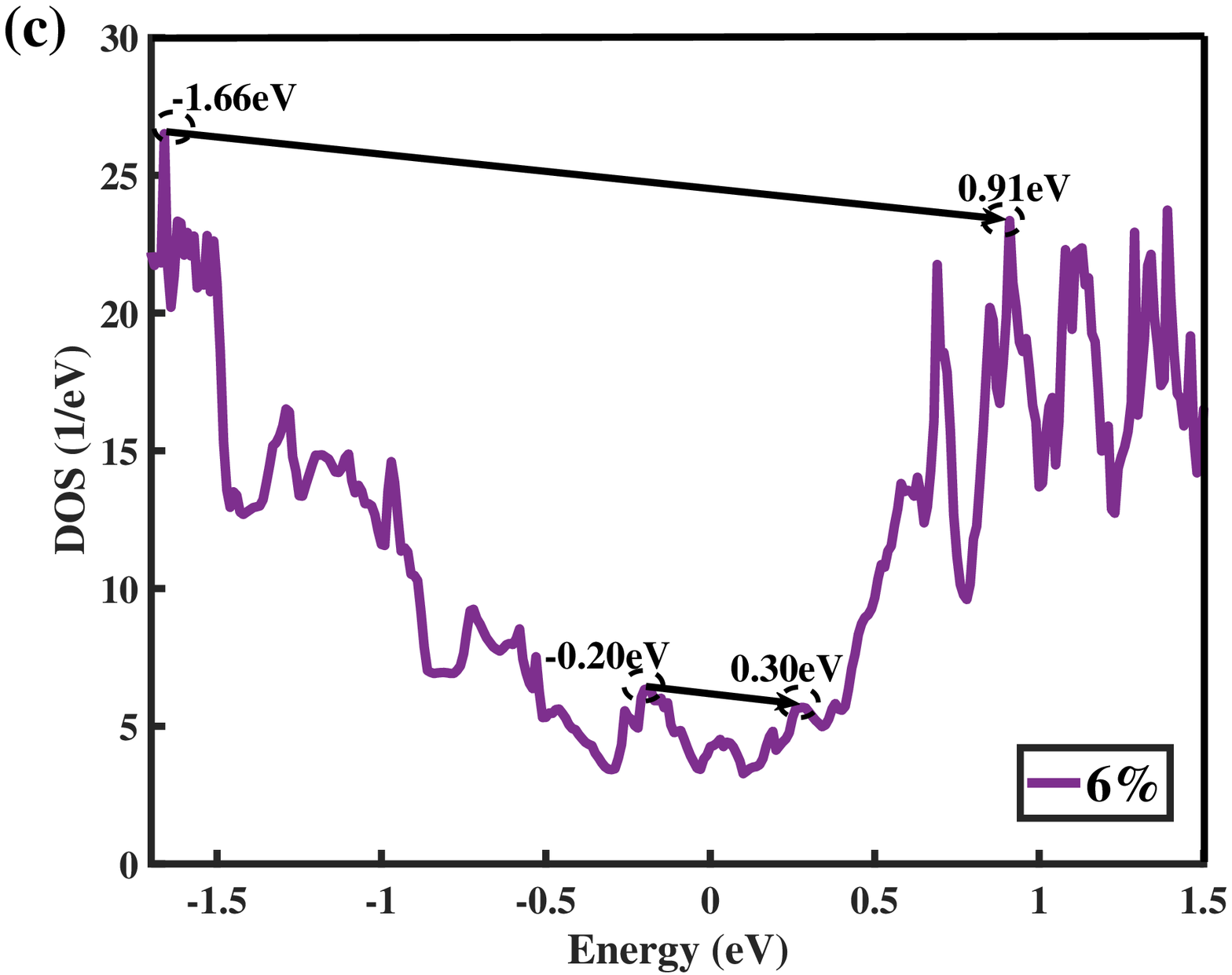}%
    \includegraphics[scale=0.2]{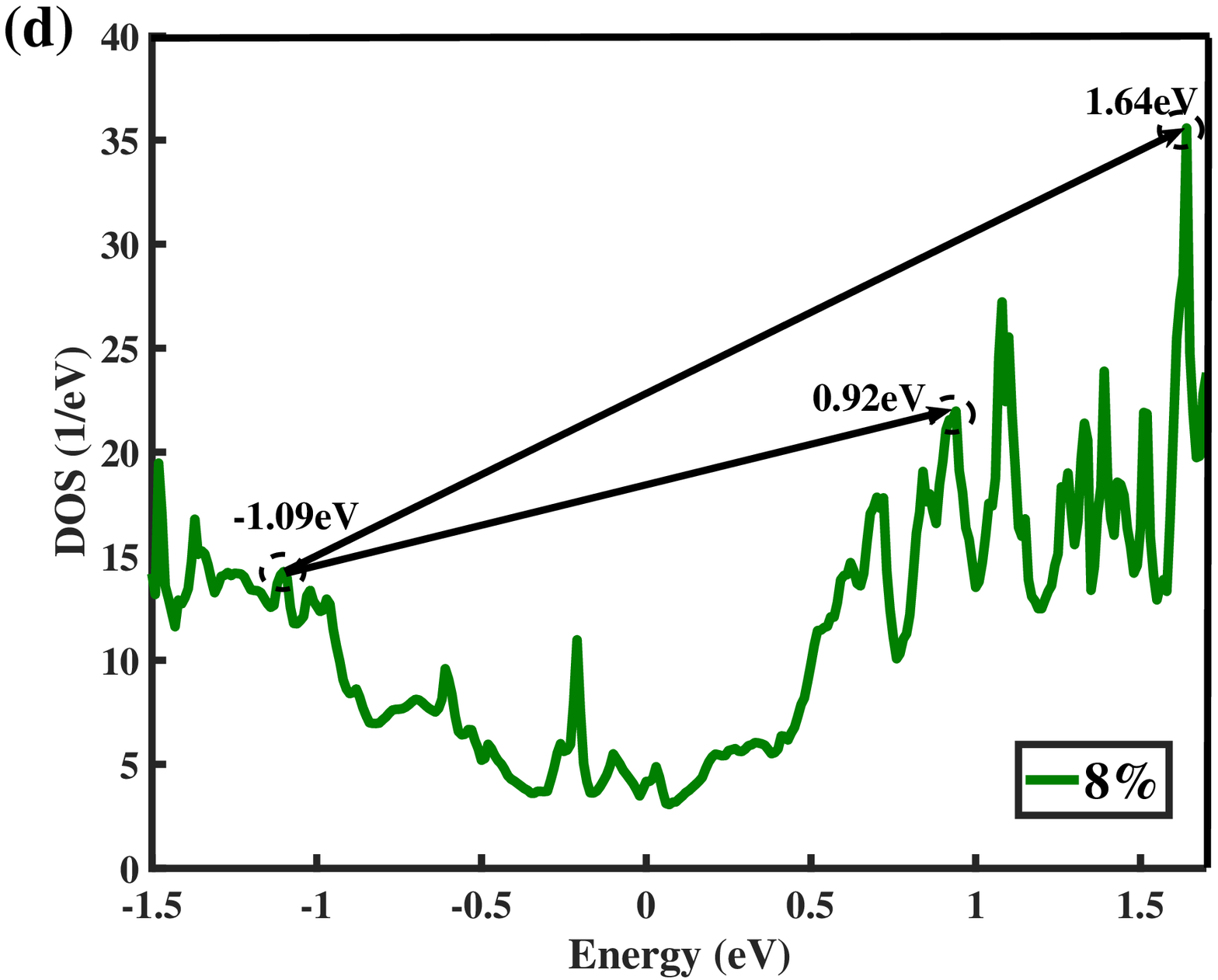}%
	\caption{Density of states of TaIrTe$_4$ for different strain strengths exerted along z direction: (a)$\zeta = 2 $, (b)$\zeta = 4\%  $ , (c)$\zeta = 6\%  $ , (d)$\zeta = 8\%$.}
	\label{fig:zDOS}
\end{figure}

Accordingly, we plot DOS of the central TaIrTe$_4$ to understand the photocurrent behaviors, as shown in Fig.~\ref{fig:zDOS}. when $\zeta=2\%$,
we can observe two transitions between peaks of DOS [see black arrows], one is the transition from -0.01eV to 0.29eV corresponding to the photocurrent peak at 0.3eV, the other is the transition from -0.18eV to 0.71eV coinciding with the large photocurrent at 0.9eV, as shown in Fig.~\ref{fig:zphc}. Similarly, when $\zeta=4\%$, two transitions are plotted to describe the two photocurrent peaks at the photon energies of 0.3eV and 2.5eV.
With the strain increasing to $\zeta=6\%$, the photocurrent peaks at 0.5eV and 2.6eV are in accord with two transitions given in Fig.~\ref{fig:zDOS}(c).
Certainly, there also exist two transitions between peaks of DOS to describe the photocurrent peaks occurring  at the photon energies of 2.0eV and 2.7eV for $\zeta=8\%$. Similarly, the photocurrent can be well explained in terms of the transitions between DOS.

In summary, we investigate the effect of the strain on the energy bands of TaIrTe$_4$ sheet and the photocurrent in the Cu-TaIrTe$_4$-Cu heterojunction.
It is found that the Weyl points remain intact and do not separate for the strain exerted along x direction. The band gap near the hight-symmetry point has a relative shift to become an indirect band gap. The maximum photocurrent is acquired at either $\theta=0^{\circ}$ or $\theta=90^{\circ}$ for all photon energies.
The maximum photocurrent shows that one can obtain a large photocurrent in the Cu-TaIrTe$_4$-Cu heterojunction in the absence of the strain.
While the photocurrent can be sharply enhanced and reach more than 86 times of the photocurrent at the case of zero strain. Accordingly, the maximum values of the photocurrent can be explained in terms of the transitions between DOS peaks and band structures. The strain-induced energy bands and photocurrent exhibit anisotropic behaviors. When the strain is exerted along z direction, the Weyl points can be completely broken with increasing of the strain. In the infrared-light region, the strain can result in a significant decrease of the photocurrent. The strain can be utilized to modulate
the photocurrent and detect the variation of energy bands of Weyl semimetals. Our results shed ligths on the promising application of photogalvanic effects based on TaIrTe$_4$ devices in photonics and optoelectronics.

This work was supported by National Natural Science Foundation of China (Grant No. 11574067).

\textbf{Conflict of interest}
The authors have no conflicts to disclose.

\textbf{DATA AVAILABILITY}
The data that support the findings of this study are available from the corresponding author upon reasonable request.

\end{document}